\documentclass[twocolumn]{aastex631}
\usepackage{graphicx} 

\newcommand{\unit}[1]{{\ \rm #1}}
\newcommand{\txt}[1]{{\rm #1}}
\newcommand{\Hmol}{{\rm H_2}}

\newcommand{\m}{\ensuremath{\,{\rm m}}}

\newcommand{\K}{\ensuremath{\, {\rm K}}}

\renewcommand{\deg}{\ensuremath{\,{\rm deg}}}

\shorttitle{CSS-FAST cross correlation}
\shortauthors{Deng et al.}
\graphicspath{{./}{figures/}}

\begin{document}

\title{Forecasting the Cross-Correlation of the CSST galaxy survey with the FAST HI Intensity Map}

\author{Furen Deng}
\author{Yan Gong}
\author{Yougang Wang}
\affiliation{National Astronomical Observatories, Chinese Academy of Sciences, Beijing 100101, China}
\affiliation{School of Astronomy and Space Science, University of Chinese Academy of Sciences, Beijing 100049, China}
\author{Shutong Dong}
\affiliation{School of physics, Nankai University, Tianjin, 300071, China}
\author{Ye Cao}
\affiliation{Department of Astronomy, Beijing Normal University, Beijing 100875, China}
\author{Xuelei Chen}
\affiliation{National Astronomical Observatories, Chinese Academy of Sciences, Beijing 100101, China}
\affiliation{School of Astronomy and Space Science, University of Chinese Academy of Sciences, Beijing 100049, China}
\affiliation{Department of Physics, College of Sciences, Northeastern University, Shenyang 110819, China}
\affiliation{Center of High Energy Physics, Peking University, Beijing 100871, China}








\begin{abstract}
The cross-correlation of optical galaxies with the neutral hydrogen (HI) radiation intensity can enhance the signal-to-noise ratio (SNR) of the HI intensity measurement. In this paper, we investigate the cross-correlation of the galaxy samples obtained by the spectroscopic survey of the China Space Station Telescope (CSST) with the HI Intensity mapping (IM) survey of the Five-hundred-meter Aperture Spherical Telescope (FAST). 
Using the IllusitrisTNG simulation result at redshift $0.2 \sim 0.3$, we generate mock data of the CSST survey and a FAST L-band drift scan survey. The CSST spectroscopic survey can yield a sample of galaxies with a high comoving number density of $10^{-2} (\unit{Mpc}/h)^{-3}$ at $z \sim 0.3$.
We cross-correlate the foreground-removed radio intensity with the CSST galaxies, including both the whole sample, and red and blue galaxy sub-samples separately. We find that in all cases the HI and optical galaxies are well correlated. The total HI abundance can be measured with a high precision from this correlation. A relative error of  $\sim 0.6\%$ for $\Omega_{\rm HI}$ could be achieved at $z\sim 0.3$ for an overlapping survey area of $10000 \unit{deg}^2$.
\end{abstract}


\section{Introduction} \label{sec:intro}
 
Neutral hydrogen plays a key role in galaxy formation and evolution, as the fuel for the formation of denser structures such as molecular clouds and stars, and it is also a tracer of the large scale structure of the Universe. The redshifted 21 cm signal produced by neutral hydrogen (HI) contains rich cosmological information, which can be used for testing cosmological models and constraining cosmological parameters. Instead of observing individual galaxies one-by-one, HI intensity mapping is an efficient technique to perform fast surveys over large cosmic volumes in wide redshift range, by detecting integrated signal \citep{chang08}. 
A number of intensity mapping experiments are underway, such as the Tianlai \citep{2012IJMPS..12..256C,2015ApJ...798...40X,li2020tianlai, wu2021tianlai}, CHIME \citep{2014SPIE.9145E..22B,2014SPIE.9145E..4VN,amiri2022detection} and HIRAX \citep{2016SPIE.9906E..5XN}, BINGO \citep{2012arXiv1209.1041B,2016arXiv161006826B} experiments. This technique
may also be applied to the new generation of general purpose radio telescopes, such as the MeerKAT \citep{santos2017meerklass,wang2021h} 
and the Square Kilometer Array (SKA) \citep{2015aska.confE..19S}. 

The Five-hundred-meter Aperture Spherical Telescope (FAST) \citep{nan2011five} is a single dish telescope with currently the largest aperture (300 m during operation) and is equipped with multi-beam feed system (19 beams presently for L band) and low-noise cryogenic receivers, making it very suitable for large HI intensity mapping surveys\citep{2016ASPC..502...41B,Hu:2019okh}.

However, in intensity mapping observations a major obstacle is the foreground contamination. The Galactic synchrotron and free-free emission are at least three to four orders of magnitude brighter than the 21 cm signal at the L-band. In principle, the foreground can be removed using various algorithms, for example, log-polynomial fitting, PCA and ICA (for a review see e.g. \citealt{asorey2020hir4}). However, signal loss and foreground residual are inevitable, and additionally there are also calibration errors, radio frequency interference (RFI) and receiver noise that affect the observation. To-date, intensity mapping signal has not yet been positively detected in auto-correlations.

One method to reduce the contamination is to correlate the radio signal with the optical galaxy catalog. As the foreground is mostly Galactic and not correlated with the 21cm signal, such correlation will enhance the signal-to-noise ratio (SNR), making such cross-correlation easier to detect than the auto-correlations. 

\begin{deluxetable*}{lll}
     \tablecaption{$10^3\Omega_{\rm HI} b_{\rm HI} r_{\txt{HI}, g}$ measurement at redshift $z\approx 0.8$ in literature.}\label{tab:hi_bias_compare}
     \tablehead{
     \colhead{Survey} & \colhead{Reference} & \colhead{$10^3\Omega_{\rm HI} b_{\rm HI} r_{\txt{HI}, g}$}
     }
     \startdata
     GBT-DEEP2 &\citet{chang2010intensity}  & $0.55\pm 1.5$ \\
     GBT-WiggleZ  & \citet{masui2013measurement}& $0.43\pm 0.07(\rm stat)\pm 0.04(\rm sys)$ \\
     GBT-WiggleZ & \citet{wolz2021hi} & $0.58\pm 0.09(\rm stat)\pm 0.05(\rm sys)$ \\
     GBT-eBOSS ELG  &\citet{wolz2021hi} & $0.40\pm 0.09(\rm stat)\pm 0.04(\rm sys)$ \\
     GBT-eBOSS LRG & \citet{wolz2021hi} & $0.35\pm 0.08(\rm stat)\pm 0.03(\rm sys)$ \\
     \enddata
\end{deluxetable*}

Indeed, positive detection of the cross-correlation has been achieved by correlating the Green Bank Telescope (GBT) data with the DEEP2 survey \citep{chang2010intensity}, WiggleZ Dark Energy survey \citep{masui2013measurement}. 
More recently, the HI IM survey data from the Parkes telescope data is correlated with 2dF survey and a detection at $z\approx 0.1$ is reported \citep{Anderson2018}, though the cross-correlation is lower than expected. The GBT HI IM survey data is also correlated with the eBOSS ELG and the eBOSS LRG to constrain HI abundance and bias \citep{wolz2021hi}. Some GBT quantitative results at redshift $z\approx 0.8$ are summarized in Table.~\ref{tab:hi_bias_compare}. More recently, the Canadian Hydrogen Intensity Mapping Experiment (CHIME)
detected HI cross-correlation with the luminous red galaxies (LRG) at $z=0.84$, emission line galaxies (ELG) at $z=0.96$, and quasars (QSO) from the eBOSS clustering catalogs at $z=1.20$ by stacking the map on the angular and spectral locations of optic galaxies, though the errors in the amplitudes of correlation are still relatively large \citep{CHIME2022}. 

Besides the technical advantage, the cross-correlation of optical galaxies and HI intensity also offers useful information for science: how the neutral hydrogen and stellar mass are correlated. It is believed that on large scales both will trace the total density which is dominated by the dark matter, but on small scales the relation could be complicated, depending on a number of galaxy evolution parameters. The neutral hydrogen is linked with the current star formation rate, while the luminosity of the galaxy is determined to a large extent by the stellar mass, which has been accumulated over the entire cosmic history. By statistical analysis of the correlation between the HI and different types of optical galaxies, one may gain further insights on the formation and evolution of galaxies \citep{wolz2019intensity}.

The Chinese Space Station Telescope (CSST) has a planned Optical Survey (CSS-OS) \citep{zhan2011consideration,gong2019cosmology}, which includes both multi-band photometric survey and low resolution spectroscopic survey. The CSS-OS is planned to cover a total sky area of $17500\unit{deg^2}$. It will provide a large set of galaxies over a broad redshift range, suitable for correlation with intensity mapping data. In the present paper, we investigate the potential of cross-correlation between the FAST HI intensity mapping survey with the CSS-OS spectroscopic survey galaxies. 

In principle, the HI auto-correlation power spectrum and the HI-optical galaxy number density cross-correlation power spectrum are proportional to the matter density power spectrum, and can be used to probe many cosmological parameters. For example, the baryon acoustic oscillation (BAO) wiggles can be used to constrain the cosmological distance scale and Hubble expansion parameters. However, at present most studies of HI-optical galaxy cross-correlation have focused on the measurement of total neutral hydrogen density $\Omega_{\rm HI}$ at the redshift of the survey. This is because at present, foreground removal is still a very great challenge in any cosmological 21 cm observation, the residue error is usually far greater than the statistical error expected for the power spectrum. It is difficult to give an accurate estimate of the uncertainty in the HI auto-correlation power spectrum at the present time, even if we succeed in detecting it in the future survey. As for the HI-optical cross-correlation power spectrum, we are more confident of its successful measurement, which has been demonstrated by the GBT, Parkes and CHIME experiments, but by definition we would already have the optical galaxy measurement in place, which would provide a good constraint on the general cosmological parameters, it is not expected that the HI-optical cross-correlation power could give much improvement over the optical measurement. The total neutral hydrogen density $\Omega_{\rm HI}$ can however be measured directly from the HI cross power spectrum, and it is not available from the optical observations, at least not directly. While $\Omega_{\rm HI}$ is not a fundamental parameter in the standard cosmological model, it is a physical parameter of the low redshift Universe and is of great interest to the galaxy evolution model. We shall also focus on this parameter in the present paper. 

The structure of this paper is as follows. 
In Section~\ref{sec:TNG_simu}, we discuss how to generate mock observational data from the simulation. The construction of HI catalog and optical galaxy catalog are introduced in Sec.~\ref{sec:higal} and Sec.~\ref{sec:gal_cat}, the simulation of intensity mapping is described in Sec.~\ref{sec:IM}, and the foreground removal and transfer function calculation for the cross-correlation power spectrum are presented in Section~\ref{sec:fgrm}. 
In Section~\ref{sec:power}, we calculate the auto-correlation and cross-correlation power spectra, and forecast the precision of cosmological parameters. We summarize and conclude in Section \ref{sec:conclusion}.

\section{Simulation} \label{sec:TNG_simu}
We use a modern cosmological simulation, the IllustrisTNG \citep{pillepich2018simulating}, to study the expected optical-21cm cross-correlation. The IllustrisTNG is a set of state-of-the-art cosmological magnetohydrodynamical simulations completed with the moving mesh code AREPO \citep{springel2010pur}. 
It is built upon the Illustris galaxy formation model, and includes physical processes such as gas cooling, star formation, stellar evolution, metal enrichment, black hole growth, stellar winds, supernovae, and active galactic nuclei (AGNs) \citep{weinberger2016simulating,pillepich2018simulating}. 
It has been compared with observations and found to be in good agreement (see, e.g. \citealt{nelson2018first,diemer2018modeling}). In this work, we use the TNG300-1 data set, the corresponding box size is $205\unit{Mpc}/h$, with $2500^3$ resolved elements for both dark matter and baryon. 
Some important simulation parameters are listed in Table \ref{tab:TNG_details}. 
The IllustrisTNG adopted the Planck 2015 cosmological parameters \citep{ade2016planck}, i.e., $\Omega_\Lambda=0.6911$, $\Omega_m=0.3089$, $\Omega_b=0.0486$, $\sigma_8=0.8159$, $n_s=0.9667$, and $h=0.6774$.  The neutral hydrogen mass for each gas particle in the simulation is given directly in the TNG dataset, but the relative abundance of atomic and molecular hydrogen (HI and $\txt{H_2}$ hereafter) are not given, which we shall model as described below.  

 In this paper we are mainly interested in the neutral hydrogen at low redshift, which can be effectively covered by the L-band multi-beam receiver of FAST (1050 to 1450 MHz, corresponding to redshift up to $z\approx 0.35$ for HI survey.). We will build our HI and optical galaxy catalog by using the simulation data from snapshot 84 (hereafter snap84) with central redshift $z\approx 0.2$, and snapshot 78 (hereafter snap78) with $z\approx 0.3$. 

\subsection{The {\rm HI} model}\label{sec:higal}

We construct a catalog of galaxies from the IllustrisTNG simulation. 
The total mass of atomic and molecule hydrogen is directly given in the simulation for each gas particle, but the two parts are not individually given. The molecular part makes a substantial fraction (e.g. \citet{diemer2018modeling,lagos2015molecular}), so we need to model the atomic-to-molecular transition and separate the two parts. 
\citet{diemer2018modeling} has constructed atomic and molecular hydrogen catalog for galaxies in IllusitrisTNG using five different models for the atomic-to-molecular transition, but their results are only available for snapshot with redshift $z=0$ and $z\geq 0.5$, which does not cover the redshift range for FAST. Following \citet{lagos2015molecular}, we model the $\Hmol$ abundance $f_{\rm mol}$  using the ratio of $\Hmol$ surface density to total neutral hydrogen surface density, $ f_{\rm mol} \equiv \Sigma_{\rm{H_2}}/(\Sigma_{\rm HI+H_2})$,
with the $\Hmol$ model from \citet[hereafter GK11]{gnedin2011environmental},
\begin{eqnarray}
     \label{eq:GK11_frac}
     f_{\rm{mol}} &= \left (1+\frac{\Sigma_c}{\Sigma_{\rm HI+H_2}}\right)^{-2},
\end{eqnarray}
where $\Sigma_c$ is the critical threshold surface density given by:
\begin{eqnarray}
     \label{GK11_crit}
    \Sigma_c &=& 2\times10^7\frac{\txt{M_\odot}}{\rm{kpc}^2}
    \frac{\left[\ln(1+gD^{3/7}_{\rm{MW}}(U_{\rm{MW}}/15)^{4/7})\right]^{4/7}}{D_{\rm{MW}}\sqrt{1+U_{\rm{MW}}D_{\rm{MW}}^2}}
\end{eqnarray}
where $D_{\rm MW}$ and $U_{\rm MW}$ are the dust-to-gas ratio and the ultraviolet (UV) field strength respectively, both in units of the local value. The factor $g$ is given by 
\begin{equation}
g = (1+\alpha s+s^2)/(1+s),
\end{equation}
where:
\begin{eqnarray}
     \alpha &=& 5\frac{U_{\rm{MW}}/2}{1+(U_{\rm{MW}}/2)^2},\\
     s &=& \frac{0.04}{D_{\rm{MW}}+1.5\times10^{-3}\times\ln(1+(3U_{\rm{MW}})^{1.7})}.
\end{eqnarray}
The dust-to-gas ratio is assumed to be proportional to the gas metallicity, thus 
\begin{equation}
D_{\rm MW}=Z_{\rm MW}=Z/Z_\odot, 
\end{equation}
where $Z_\odot=0.0127$ is the solar metallicity. For star forming cells, the UV field strength is assumed to be proportional to the star formation rate surface density, 
\begin{equation}
U_{\rm MW} = \Sigma_\txt{SFR}/\Sigma_\txt{SFR, local}.
\end{equation}
For quiescent cells the UV field is set to be the background value. 

\begin{table}
\caption{Simulation parameters for TNG300. 
}\label{tab:TNG_details}
\begin{tabular}{c|l}
\hline\hline
Name & Value \\
\hline
$L_{\txt{box}}$ & $205\unit{Mpc/h}$\\
$N_{\txt{gas}}$ & $2500^3$\\
$N_{\txt{DM}}$  & $2500^3$\\
$N_{\txt{TRACER}}$ & $2500^3$\\
$m_{\txt{baryon}}$ & $1.1\times 10^7\unit{M_{\odot}}$\\
$m_{\txt{DM}}$ & $5.9\times 10^7\unit{M_{\odot}}$\\
\hline\hline
\end{tabular}
\end{table}

There are a number of way to define $U_{\rm MW}$ for the background, see e.g. \citet{diemer2018modeling}. We define the background by comparing the flux at $1000\ \textup{\AA}$ as \citet{gnedin2011environmental}. 
The UV background is from the 2011 updated version of \citet{faucher2009new}, which is redshift dependant. 
The local UV flux value  is   $F=3.43\times 10^{-8}\ {\rm photons}\unit{s^{-1} cm^{-2} Hz^{-1}}$ at $1000\ \textup{\AA}$
\citep{draine1978photoelectric}.

To convert the volume density in the simulation to surface density, we approximate the size of the self-gravitating system by its Jeans length $\lambda_J$ (cf. e.g. \citealt{schaye2001model}), the surface density of quantity $X$ is 
\begin{eqnarray}
     \Sigma_X &=& \rho_X\lambda_J, \quad   \lambda_J = \sqrt{\frac{\gamma(\gamma-1)u}{G\rho_\txt{gas}}}.
\end{eqnarray}
where $\rho_X$ is the volume density, $ \rho_\txt{gas}$ is the gas density, 
$u$ the internal energy per unit mass, and $\gamma=5/3$ is the adiabatic index. 
Although the Jeans length approximation exhibits large cell-to-cell scatter, the scatter is comparable with the scatter between different $\Hmol$ models, so this would not cause too much error, and it is computationally efficient \citep{diemer2018modeling}. All qualities are calculated cell-by-cell.

\begin{figure}
\gridline{
\fig{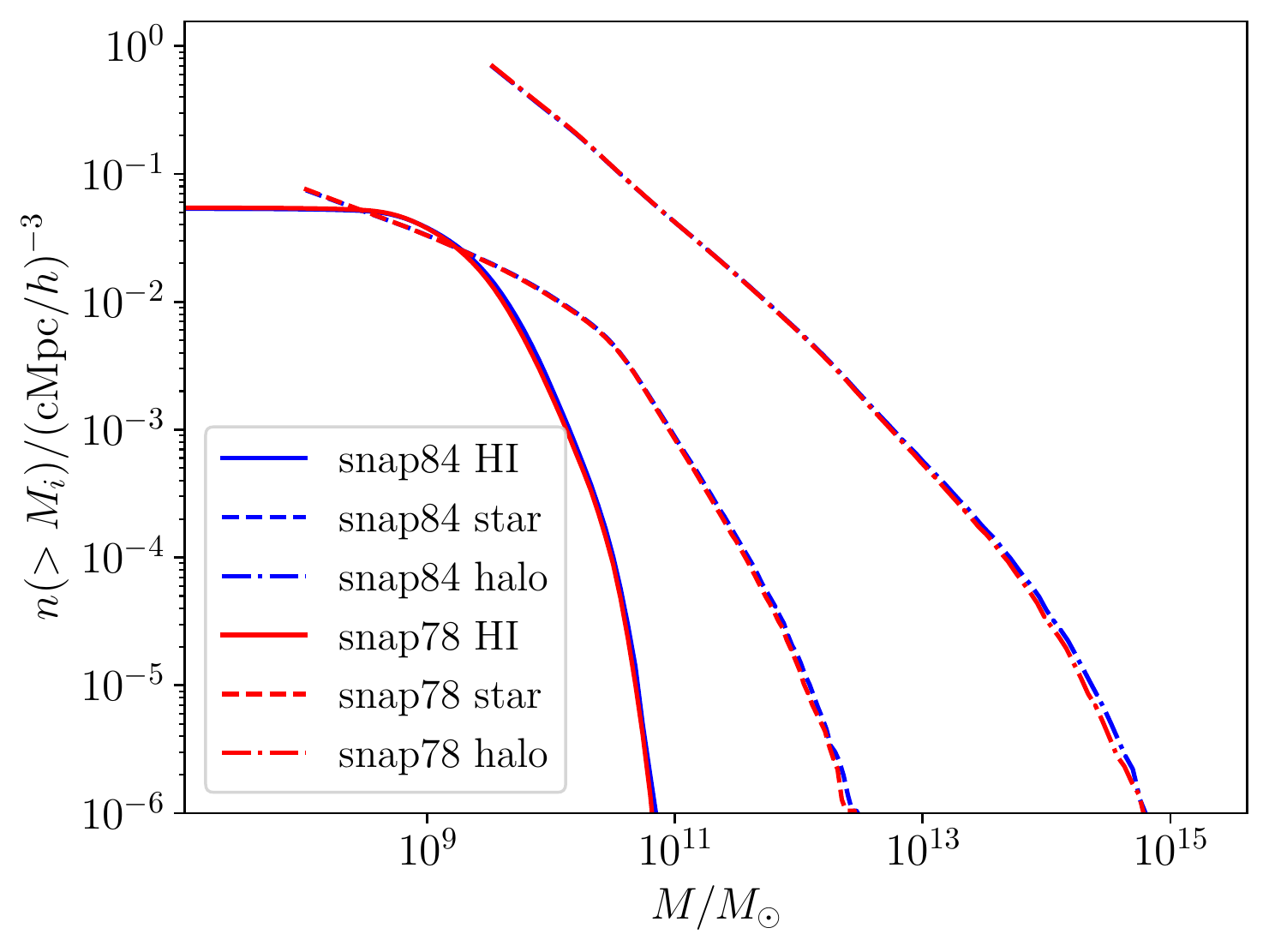}{0.45\textwidth}{}
        }
\caption{Cumulative galaxy HI mass function, stellar mass function, and halo mass function for the two snapshots snap84 and snap78. The left y-axis is for stellar and HI with $i={\rm HI,\,star}$ and the right y-axis is for halo.
}\label{fig:himass_func}
\end{figure}

The HI mass function and stellar function of galaxies, and the halo mass (defined as the total mass within a sphere whose mean density is 200 times the mean density of the Universe at the time of the snapshot.) function for snap84 and snap78 are shown in Fig~\ref{fig:himass_func}. 
Given the mass resolution of the TNG300 simulation\citep{diemer2018modeling}, in plotting the mass function we select galaxies with stellar mass larger than $5\times 10^{10} M_\odot$ or gas mass larger than $5\times 10^9 M_\odot$ for HI mass function, as a result below that mass our HI mass function is not accurate.  However, we keep all HI ingredients, including those with smaller masses, for the intensity mapping calculation. We cut the stellar mass function at $10^8\ M_\odot$ and the halo mass function at $3\times 10^9 M_\odot$ (50 dark matter particles in TNG300).
As shown by Fig.~\ref{fig:himass_func}, there is little evolution between the two snapshots.  

\subsection{The CSS-OS Galaxy catalog}\label{sec:gal_cat}  
The CSST is equipped with seven photometric band filters, and slitless grating spectrographs for three bands covering $255-1000\unit{nm}$. The angular resolution is $\sim 1.5'$ for photometric imaging, and $\sim 0.3'$ for spectroscopy (80\% energy concentration radius). The spectroscopic resolution $R=\lambda/\Delta\lambda$ is better than 200 (see Tab.~\ref{tab:spec_params}). The CSS-OS will include both a photometric imaging survey and a spectroscopic survey, covering a wide field of  $17500\unit{deg^2}$ 
in about ten years. 
The photometric catalog will be used to select targets for the spectroscopic survey. 
The photometric redshift has too large uncertainty to be of much use for cross-correlation, below we will consider the 
cross-correlation between the HI survey and the CSS-OS spectroscopic survey. 

First we consider the construction of galaxy catalog for the CSS-OS photometric and spectroscopic surveys. The star formation process is already included in the physical model of TNG. 
However, to obtain the galaxies that can be detected by the CSST spectrometer and photometer, we need to calculate the spectrum for each galaxy, and convolve it with the instrument bandpass.  We adopted model B in \citet{nelson2018first}, which matches observation reasonably well regarding the color of galaxies. The spectrum for each star particle is calculated by the FSPS stellar population synthesis code \citep{conroy2009propagation,conroy2010propagation,foreman2014python}, which models it as a single-burst simple stellar population (SSP). 
We adopted the Padova isochrones (\citep{padovaI,padovaII}), MILES stellar library (\citet{miles06,miles11}), and assuming a Chabrier initial mass function (\citet{Chabrier_2003}) following \citet{nelson2018first}. 
Cosmological parameters are also set to TNG values.

The host galaxy extinction is modeled as a power law with two dust components as described in \cite{charlot2000simple}:
\begin{eqnarray}\label{eq:powerlaw_ext}
    \tau_\lambda &=& 
    \left\{
    \begin{array}{cc}
    \tau_1 (\lambda/\lambda_0)^{-\alpha_1}, & t_\txt{ssp}\leq t_\txt{age}\\
    \tau_2 (\lambda/\lambda_0)^{-\alpha_2}, & t_\txt{ssp}> t_\txt{age}\\
    \end{array}
    \right.,
\end{eqnarray}
where $\tau_1=1.0$, $\tau_2=0.3$, $\alpha_1=\alpha_2=0.7$, $t_\txt{age}=10\unit{Myr}$ and $\lambda_0=550\unit{nm}$. 
The rest-frame spectrum of a star particle is attenuated by $L_\txt{rest}(\lambda)=L_i(\lambda)\exp(-\tau_\lambda)$. 

The CSST photometer is equipped seven filters: NUV, u, g, r, i, z and y, and the spectrometer is equipped with three filters: GU, GV and GI. 
For photometer, the AB magnitude is given by: 
\begin{eqnarray}\label{eq:obs_mag}
m_{\rm AB} &=& -2.5\log_{10}\left( \frac{(z+1)}{4\pi r_L^2 }  \frac{\int L R_X ~\mathrm{d}\ln\nu }{ \int R_X ~\mathrm{d}\ln\nu }\right) - 48.60,\nonumber\\
\end{eqnarray}
where $L(\nu)= L_\txt{rest}[(1+z)(1+\frac{v_\txt{los}}{c})\nu] $ is the galaxy luminosity, which is assumed to be proportional stellar mass, $v_{\rm los}$ is the line-of-sight (LOS) component of peculiar velocity, $r_L$ is the luminosity distance, and $R_X(\nu)$ is the throughput for band $X$.

\begin{figure}
\fig{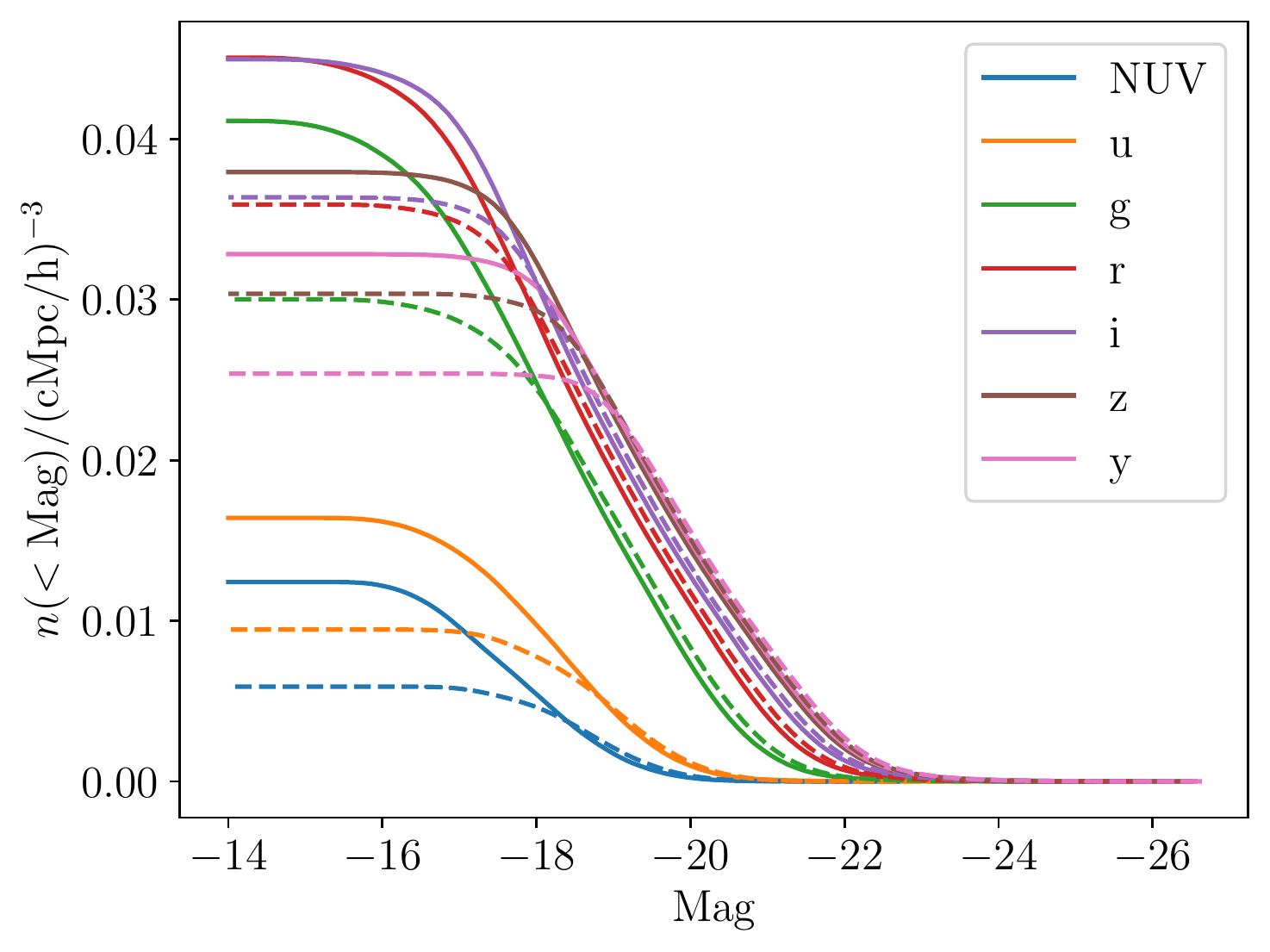}{0.45\textwidth}{}
\caption{
The expected CSST galaxy luminosity function ($\mathrm{SNR} > 10$)  for each photometry band. The solid line is for snap84 and the dashed line is for snap78.
\label{fig:lum_func}
}
\end{figure}

Precise measurement of the redshift of a galaxy depends on its emission line or absorption line features. Thus, to obtain the galaxies that can be detected by the CSST spectrometer, we choose six strong atomic lines 
$\txt{H}\alpha$, $\txt{H}\beta$, $\txt{OIII}$, $\txt{H}\delta$, $\txt{CaK}$, $\txt{CaH}+\txt{H}\epsilon$, 
as indicators for whether a galaxy can be detected by CSST spectrometer. The first three lines are mainly detected as emission line features, while the last three are mainly detected as absorption feature. For each feature we measure the flux within a window, 
and we also measure the flux of the continuum around each windows, to obtain the contrast of the emission or absorption feature. The flux contrast and continuum detected by the CSST spectrometer are given by:
\begin{eqnarray}
\label{eq:dfline}
\Delta F_\txt{line}^i &=& \frac{1}{4\pi r_L^2}\int_{\nu_i-\frac{\Delta\nu}{2}}^{\nu_i+\frac{\Delta\nu}{2}}(L_\txt{rest}(\nu) - L_\txt{rest}^\txt{cont}(\nu))d\nu,\\
\label{eq:fcont}
F_\txt{cont}(\nu) &=& \frac{z+1}{4\pi r_L^2}L_\txt{rest}^\txt{cont}((z+1)(1+\frac{v_\txt{los}}{c})\nu)
\end{eqnarray}
where $\nu_i$ is the rest-frame wavelength for line $i$, $\Delta\nu$ is the size of the window, and $L_\txt{rest}^\txt{cont}$ is the continuum luminosity. We have assumed that both the emission and absorption feature are narrow enough to be fully covered by one CSST spectrometer resolution unit.

The extinction of the Milky Way is included with the extinction law $A(\lambda)/A(V)$ given in \citet{cardelli1989relationship},
and differential extinction $E(B-V)$ is from the \citet{schlegel1998maps} dust reddening map. For photometry, the extinction is calculated using the effective wavelength for each band, as
$$\nu_{\txt{eff}} = \frac{\int \nu R(\nu)\ln\nu}{\int R(\nu)\ln\nu},
\qquad
\lambda_{\txt{eff}}=\frac{c}{\nu_{\txt{eff}}}.
$$
For spectroscopy, the extinction is calculated at the observed wavelength of each spectral line. 

\begin{deluxetable}{cccc}
     \tablecaption{Wavelength range and spectral resolution for central wavelength for CSST spectrometer
     }\label{tab:spec_params}
     \tablewidth{0pt}
     \tablehead{
     \colhead{Filter} & \colhead{Wavelength/\AA} & \colhead{Central wavelength/\AA} & \colhead{Resolution}\\
     }
     \startdata
     GU & 2550-4100 & 3370 & 287\\
     GV & 4000-6400 & 5250 & 232\\
     GI & 6200-10000 & 8100 & 207\\
     \enddata
 \end{deluxetable}

We estimate the signal-to-noise ratio (SNR) for each galaxy in the photometric surveys by making circular aperture photometry for each galaxy. For the slitless spectrometer, we model each galaxy as a point source for simplicity, though we caution that at low redshift the galaxies are generally extended sources. The SNR for each photometric aperture and spectrometer sample are 
calculated as the same as in \citet{cao2018testing}: 
\begin{eqnarray}
     \txt{SNR} = \frac{|\Delta C_s| t_\txt{exp}\sqrt{N_\txt{obs}}}{\sqrt{C_s t_\txt{exp} + N_\txt{pix}[(B_\txt{sky}+B_\txt{det})t_\txt{exp} + R_\txt{n}^2}]}~~
\end{eqnarray}
where $N_\txt{pix}=\Delta A/l_p^2$, is the number of pixels occupied by the aperture, 
$l_p=0.074\unit{arcsec}$ is the pixel size. 
$B_\txt{det}=0.02\ e^-\txt{s^{-1}pixel^{-1}}$ 
is the detector dark current, $t_\txt{exp}$ is exposure time and $N_\txt{obs}$ is the number observations. 
We set $t_\txt{exp}=150\unit{s}$ and $N_{\rm obs}=4$ for spectrometer and NUV and y bands of photometer, and $t_\txt{exp}=150\unit{s}$ and $N_{\rm obs}=2$ for other photometry band (\citet{gong2019cosmology}).  
$B_\txt{sky}$ is the count rate from sky background per pixel:
\begin{eqnarray}
    B_\txt{sky} &=& A_\txt{eff}\int I_\txt{sky}(\nu)R_X(\nu)l_p^2 \frac{d\nu}{h\nu},
\end{eqnarray}
where $A_\txt{eff}=3.14 \unit{m^2}$ is the effective area for CSST, and $I_\txt{sky}(\nu)$ is the average sky background, including earthshine and zodiacal light  \citep{ubeda2011acs}, $C_s$ is the count rate from the galaxy, $\Delta C_s$ is the signal contrast. For photometer, 
\begin{eqnarray}
    \Delta C_s = C_s &=& A_\txt{eff}10^{-0.4(m_\txt{AB}+48.60)}\int R_X(\nu)\frac{d\nu}{h\nu}, 
\end{eqnarray}
and for spectrometer and spectral line $i$, 
\begin{eqnarray}
    \Delta C_s^i &=& A_\txt{eff}R_X\left(\frac{\nu_i}{z+1}\right)\frac{\Delta F_\txt{line}^i}{h\nu_i/(z+1)}, \\
    C_s^i &=& \Delta C_s + A_\txt{eff}\int_{\nu_o^i-\frac{\nu_R}{2}}^{\nu_o^i+\frac{\nu_R}{2}} R_X F_\txt{cont}\frac{1}{h\nu}d\nu,
\end{eqnarray}
where $\Delta F_\txt{line}$ and $F_\txt{cont}$ are given in (\ref{eq:dfline}) and (\ref{eq:fcont}), $\nu_o^i$ is the observed frequency for line $i$ and $\nu_R$ is the spectral resolution of CSST spectrometer. The wavelength range and resolution for central wavelength for CSST spectrometer are given in Table \ref{tab:spec_params}.

\begin{figure}
\gridline{
\fig{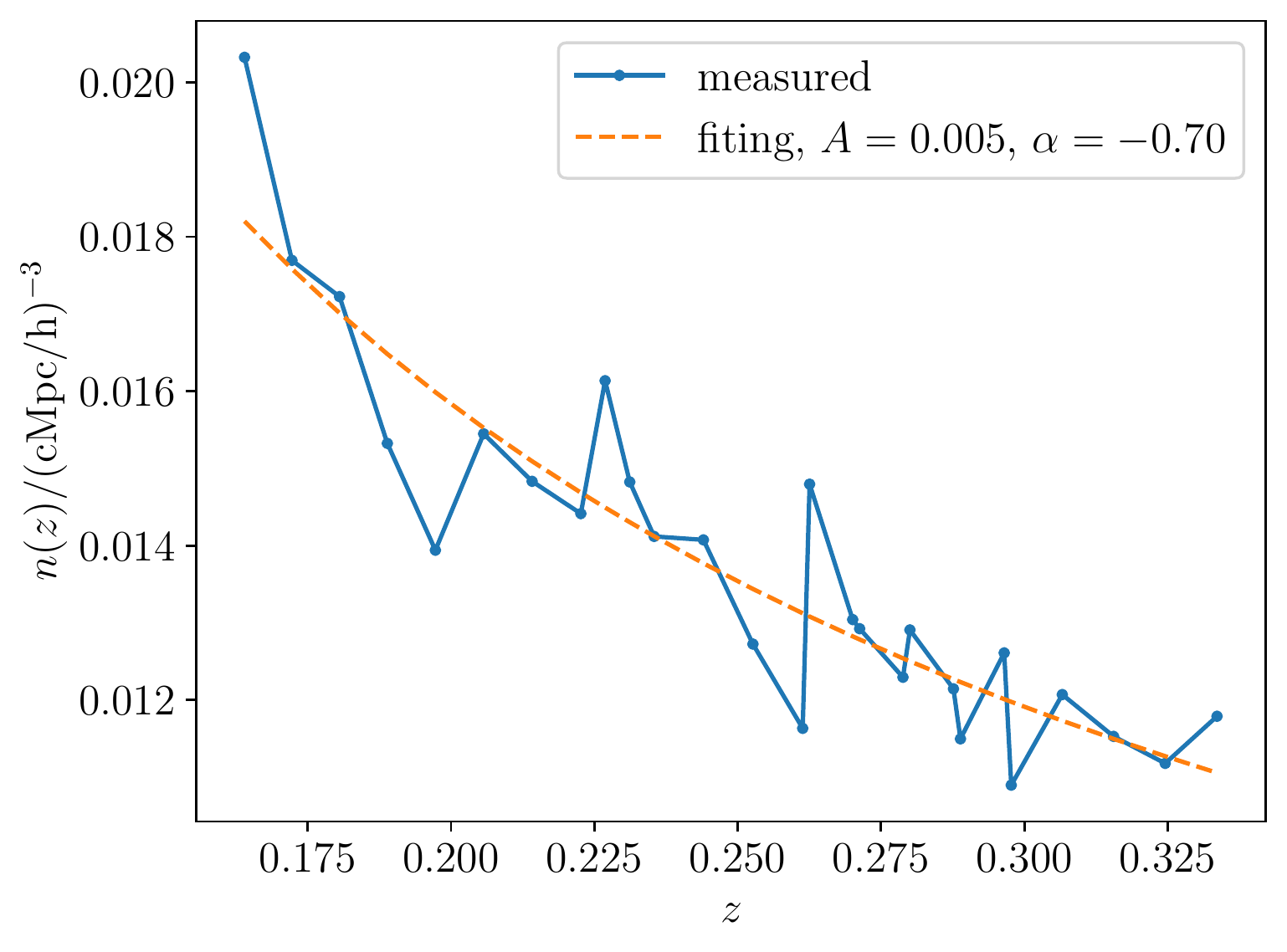}{0.45\textwidth}{}
    }
\caption{The number density of galaxies detected by the spectrometer, as a function of redshift.\label{fig:gal_distri}
}
\end{figure}

The 80\% energy concentration radius of the point spread function (PSF) is $R_{80}^\txt{spec}\approx 0.3\ \unit{arcsec}$ for the CSST spectrometer, and $R_{80}^\txt{phot}\approx 0.15\unit{arcsec}$ for the CSST photometer. 
We model the PSF of the photometer as a gaussian function, with $\sigma_\txt{PSF}^\txt{phot}\approx 0.08\unit{arcsec}$. 
For photometry, the image of each galaxy within $5\sigma_\txt{PSF}^\txt{phot}$ is then convolved with the PSF before the aperture photometry, which is sufficient to model the effect of PSF. 
For the spectrometer, the light from one aperture is spread into a spectrum. The resolution units can be estimated as a rectangular with side length $R_{80}^\txt{spec}$ along the dispersion direction and $2R_{80}^\txt{spec}$ perpendicular to the dispersion direction.

We set the detection threshold as ${\rm SNR}=10$ for the photometric catalog and 5 for the spectroscopic catalog. 
The luminosity functions of galaxy samples for each CSST photometry band are shown in Figure \ref{fig:lum_func}. 
Because the spectroscopic target is always selected from high SNR photometric catalog, when we talk about the spectroscopic catalog, the SNR should always exceed 10 for at least one photometric band. Nevertheless, we find that almost all galaxies detected by the spectrometer with SNR larger than 5 are also detected by at least one photometric band with SNR exceeding 10. 

The number density of galaxies detected by at least one spectroscopic band and at least one photometric band is shown in Figure \ref{fig:gal_distri}.
Compared with previous surveys, the galaxy number density of the CSS-OS is significantly greater. For example, the SDSS-II (main galaxy sample for DR7) is  $\sim 2\times 10^{-4} (\txt{Mpc}/h)^{-3}$  at $z\approx 0.2$ \citep{ross2015SDSSDR7}, and for BOSS (DR12) it is $\sim 5\times 10^{-4}\ \txt{to} \ 3\times 10^{-4}\ (\txt{Mpc}/h)^{-3}$ from $z=0.2$ to $z=0.3$ \citep{alamBOSSDR12}. The ongoing DESI survey is expected to have a galaxy number density of $ \sim 9\times 10^{-4}\ (\txt{Mpc}/h)^{-3}$ at $z\approx 0.3$ \citep{desi_snowmass13}, which is still significantly lower than the galaxy number density of the CSS-OS spectroscopic survey.  
The shot noise of power spectrum measurement would be greatly suppressed with the CSS-OS data. 

On the other hand, the FAST HI galaxy survey can achieve a number density of $1\times 10^{-3}\ \txt{to} \ 2\times 10^{-4}\ (\txt{Mpc}/h)^{-3}$ from $z=0.2$ to $z=0.3$ for an integration time of 384 s per beam \citep{Hu:2019okh} in the ideal case for $5\sigma$ detected galaxies, comparable to the current optical surveys, but still much lower than that of the CSS-OS.

\begin{figure}
\gridline{
\fig{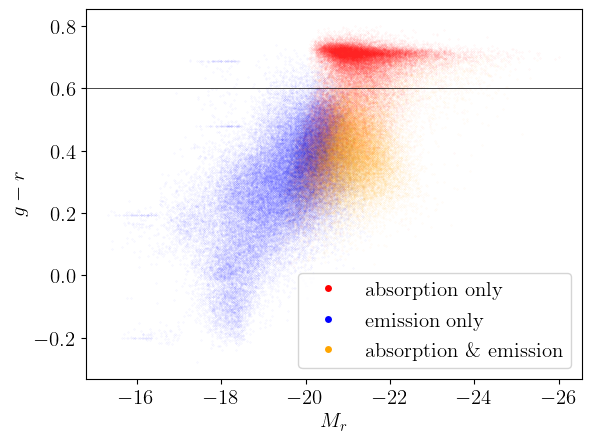}{0.45\textwidth}{}
               }
\caption{The color-Magnitude diagram for galaxies detected by only emission line, only absorption line or both by CSST spectrometer. The galaxies are detected by at least one spectroscopic band and both $g$ and $r$ bands. \label{fig:CMD_absem}
     }
\end{figure}

We select galaxies that are detected in both the spectrometer and the $r$ and $g$ photometric bands, and show them in Figure \ref{fig:CMD_absem}. Each galaxy is shown as a point, with colors to mark their spectral characteristics: those detected only in absorption spectra are colored red, those detected only in emission spectra are colored blue, while those detected in both absorption and emission spectra are colored orange. As expected, the samples of detected emission line galaxies have lower absolute luminosity (larger absolute magnitude). The color of the galaxies detected only in absorption spectra are also generally redder. 
There are a few horizontal line structures in Figure.~\ref{fig:CMD_absem}, which are artifacts from the limited time and mass resolutions of the simulation snapshots: these are some metal-poor galaxies whose stellar components are formed recently in the same time step of the simulation, but with different stellar mass. As a result, they have nearly the same colors but different absolute magnitudes. The number of such galaxies is not large in our catalog (less than 1\% of the whole catalog) and will not have significant influence on our result. 
The optical galaxy catalog can be split into two groups by $g-r= 0.6$ as marked by the black line in the plot, corresponding to blue and red galaxies.

\subsection{Intensity mapping}\label{sec:IM}
To simulate the intensity mapping observation process, the TNG300-1 box is divided into $2048\times2048\times128$ voxels.
The angular resolution is approximately $0.6\unit{arcmin}$ for snap84 and $0.4\unit{arcmin}$ for snap78 respectively, finer than the FAST beam size at this frequency. The corresponding frequency resolution is about 0.6 MHz for snap84 and 0.5 MHz for snap78 in the frequency range of interest. The frequency resolution of the FAST raw data is usually much finer, e.g. 7.6 kHz in our own pilot survey, but this finer resolution is not needed for intensity mapping, though useful for efficient RFI flagging and also for some HI galaxy observation. It is usually re-binned to a lower frequency resolution for intensity mapping data processing.
The brightness temperature of the 21 cm signal for each voxel is given by \citep{2018ApJ...866..135V} $T_b= T_0 \frac{\rho_{\txt{HI}}}{\rho_c^0}$,  
where $\rho_{\txt{HI}}$ is the comoving density of HI and $\rho_c^0$ is the critical density at $z=0$, and 
\begin{eqnarray}\label{eq:dens2Tb}
T_0 = 189h\frac{H_0(1+z)^2}{H(z)}\unit{mK}
\end{eqnarray}

The foreground is generated using GSM2016 model \citep{zheng2017improved} of the Python package PYGDSM \citep{price2016}. We have taken a square patch centered at $\txt{RA}=12\unit{h}$, $\txt{Dec}=25^\circ 48'$ (corresponding to the zenith for the FAST location), and interpolated the foreground map to the center of each voxel.

\begin{figure*}
\gridline{
\fig{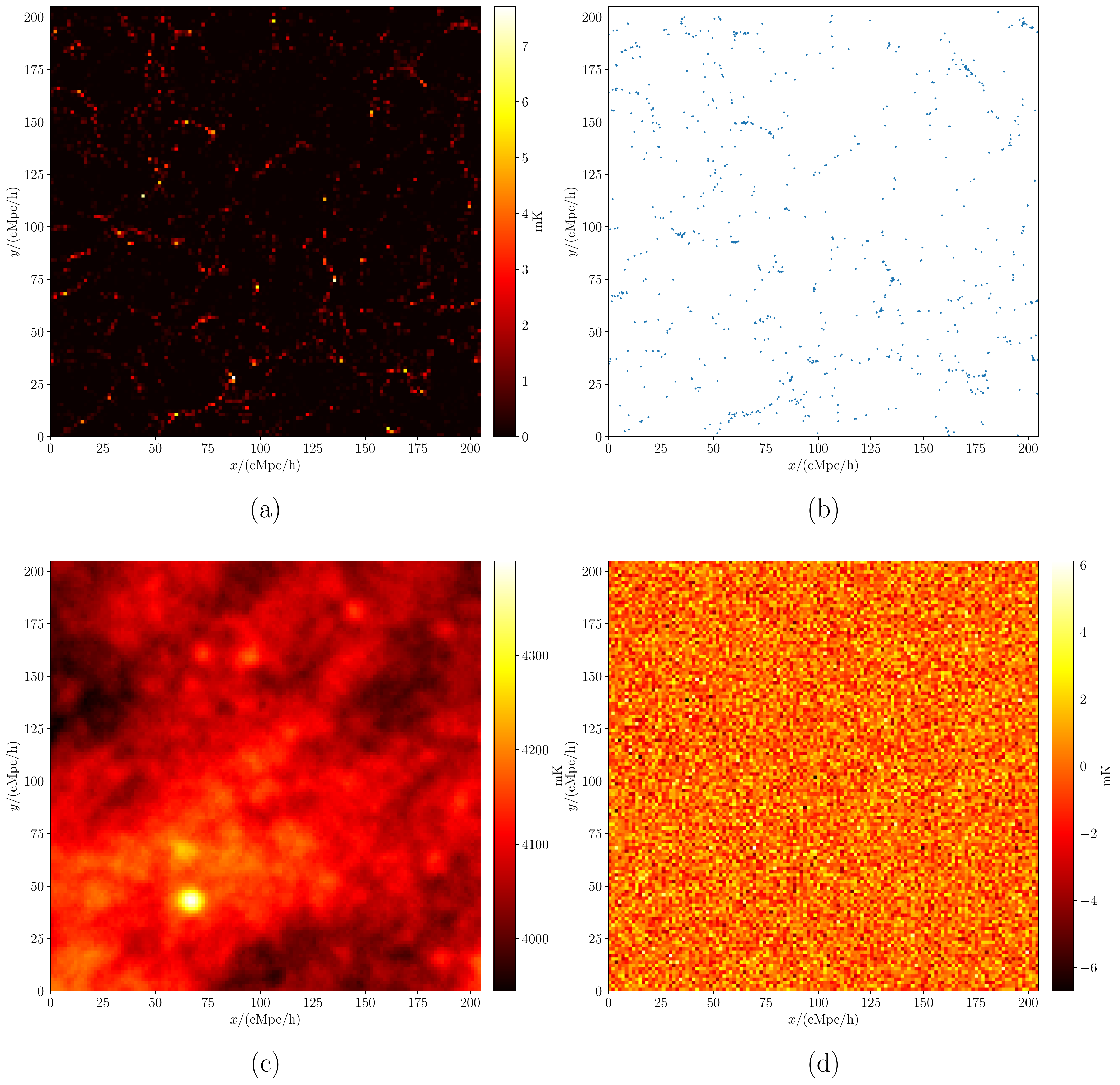}{0.95\textwidth}{}
               }
\caption{A slice at $\nu\approx 1185.725\unit{MHz}$ for (a) the 21 cm map, (b) the galaxies that can be detected by CSST spectrometer, (c) the foreground map and (d) the thermal noise map. 
\label{fig:mapmk}
     }
\end{figure*}

To simulate the drift scan survey, we generate mock time-ordered data as
\begin{eqnarray}
    T_A(x_b, y_b, \tau) &=& \int T_b(x, y)\mathcal{B}(x,y;x_b,y_b,\tau)dxdy + n_T,\nonumber
\end{eqnarray}
where $x_b$ and $y_b$ is the direction of the beam center as given in \citet{jiang2020fundamental},  $\tau=12/\cos\delta \unit{s}$ is the integration time which corresponds to the time for the sky to drift over the beam at the declination $\delta$,  and $n_T$ is the thermal noise.  Here we model the instant beam response as a Gaussian function,
\begin{equation}\label{eq:gaus_beam}
     B(x,y;x_b,y_b) = \frac{1}{2\pi\sigma^2}
     \exp\left(-\frac{(x-x_b)^2 + (y-y_b)^2}{2\sigma^2}\right),\nonumber
\end{equation}
where $\sigma = 0.518 \lambda/(300 \unit{m})$, though the real beam may have a more complicated dependence on the frequency. The averaged beam over integration time is denoted as $\mathcal{B}(x,y;x_b,y_b,\tau)$.

We model the thermal noise $n_T$ as gaussian white noise with standard deviation given by \citep{2013tra..book.....W}: 
\begin{eqnarray}\label{eq:ns_sig}
     \sigma_T &=& \frac{T_\txt{sys}}{\sqrt{2\Delta\nu\tau N_\txt{scan}}},
\end{eqnarray}
where $\Delta\nu$ is the frequency channel width, $N_\txt{scan}$ is the scan times for the sky area, $T_\txt{sys}$ is the system temperature. Following \citet{hu2020forecast}, we model the system temperature as  $T_\txt{sys}= T_\txt{rec} + T_\txt{sky}$, with the receiver temperature of about 20 K, and the sky temperature off the galactic plane as 
\begin{eqnarray}
     T_\txt{sky}=\left(2.73+25.2\left(\frac{408\unit{MHz}}{\nu}\right)^{2.75}\right)\unit{K}.
\end{eqnarray}
which yields a total system temperature of about 25 K off the galactic plane, in agreement with the measurements given in \citet{jiang2020fundamental}.
The minimum variance estimator of the map is then generated by \citep{tegmark1997make}:
\begin{eqnarray}\label{eq:tod2sky}
     \hat{T}_b &= (A^T N^{-1} A)^{-1} A^T N^{-1} T_A,
\end{eqnarray}
where $A$ is the pointing matrix and $N$ is the covariance matrix of data noise, which is assumed diagonal. We make a coarser $128\times 128\times 128$ voxels map, the side length is approximately $1.6\unit{Mpc}/h$, and the angular resolution is about $9.8\unit{arcmin}$ for snap84 and $6.6\unit{arcmin}$ for snap78. We adopt a coarse grid here, which is sufficient for the problems studied here, as we are not concerned about the smaller scales. 
One slice at about $1185.725\unit{MHz}$ for 21 cm, foreground and thermal noise for $N_\txt{scan}=1$ are shown in Figure \ref{fig:mapmk}. For comparison, the galaxies at the slice that can be detected by CSST spectrometer are also shown. 
We note that the sky is dominated by the foreground, and the thermal noise has a magnitude larger than or comparable to the 21cm signal, making its detection challenging. 
We find that in our simulation box, the temperature fluctuation of 21 cm signal is about $0.36\unit{mK}$ for snap84 and $0.40\unit{mK}$ for snap78, which are much smaller than the thermal noise level of $1.47\unit{mK}$ for snap84 and $2.85\unit{mK}$ for snap78. Nevertheless, the 21 cm power spectrum increases with scale, while the thermal noise power spectrum remains flat. For $k<0.2\ (\txt{Mpc}/h)^{-1}$, the 21 cm auto power spectrum is larger than the thermal noise, but the signal-to-noise ratio is not high. It requires a few repeated scans to reduce the noise power spectrum to achieve a good signal-to-noise ratio.

\subsection{Foreground removal and transfer function}\label{sec:fgrm}
In intensity mapping observation, the foreground radiation dominates over the signal, but it could be removed by exploiting the difference in the statistical properties of the foreground and the 21cm signal. Below, we simulate the foreground removal with two methods frequently employed in 21cm data analysis: the 
principal component analysis (PCA) and the independent component analysis (ICA).
The PCA method exploits the fact that along the line of sight (LOS) the foreground is highly correlated at different frequencies while the 21 cm signal is not. Such components are associated with the principal components in the frequency-pixel covariance matrix \citep{Masui2013,Alonso2015,Sazy2015,Zuo2019}.
The ICA method exploits the fact that the foregrounds have large non-Gaussianity while the 21cm signal is nearly Gaussian \citep{Chapman2012,Wolz2014,Alonso2015,asorey2020hir4}.

However, foreground removal will still inevitably cause some loss of 21 cm signal, especially on large scales which interested us. This problem can be compensated at least partially if we know the expected amount of signal loss in the power spectrum. This loss transfer function can be estimated using the simulation.  We simulated the mock 21 cm signal with halo and HI mass following \citet{wolz2019intensity}, \citet{murray2018powerbox} and \citet{murray2014hmf}, and set $M_\txt{min}=10^{12.3}\ M_\odot/h$ as did in \citet{wolz2021hi}.  

We create 500 mock HI maps with the same box size and spatial resolution, and for each halo the HI distribution is represented by 1000 HI particles in a cored Navarro-Frank-White (cNFW) profile \citep{2017MNRAS.464.4008P}. 
Then we project the mock map to the foreground-dominant bases obtained by PCA or ICA, and subtract the projected components from the mock map to simulate the signal loss. Denoting the mock power spectrum as $\hat{P}_0^\txt{HI}(k)$, and the cross-correlation power spectrum of the mock data 
before and after the foreground removal procedure as $\hat{P}^\txt{HI}_X(k)$, 
the transfer functions can be obtained by (\citet{wolz2021hi}):
\begin{eqnarray}
     \mathcal{T}(k) &=& \frac{\langle \hat{P}_0^{\rm HI}(k)\rangle}{\langle \hat{P}_X^{\rm HI}(k)\rangle}.
\end{eqnarray}
It can also be computed for both foreground removal methods, i.e. the PCA and ICA methods. 
 
\begin{figure}
\centering
     \gridline{\fig{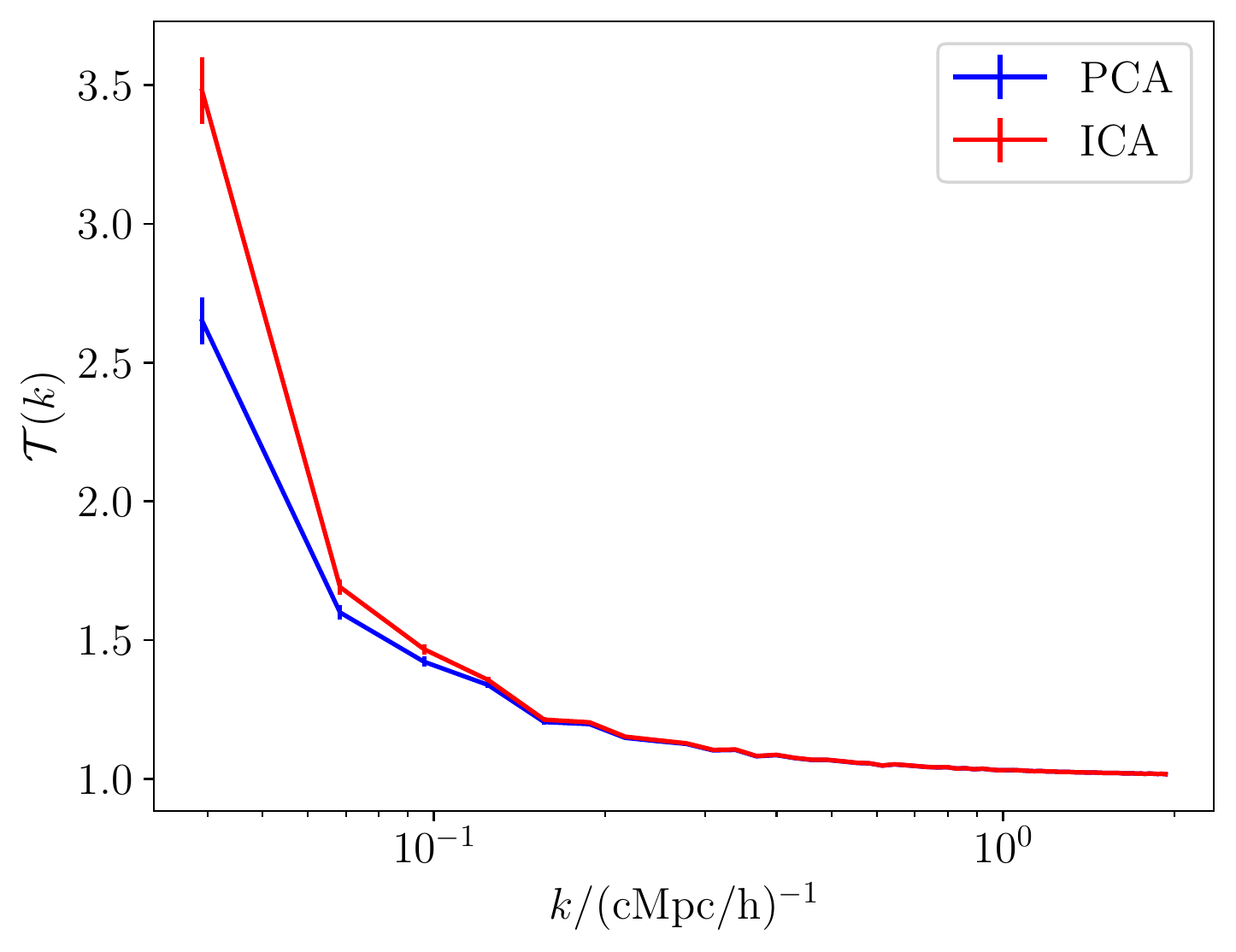}{0.45\textwidth}{}
               }
     \caption{The transfer function $\mathcal{T}_X(k)$ for both PCA and ICA, for snap84. 
     The errorbar is estimated directly from the standard deviation between mock data. 
     \label{fig:cmp_trans}}
\end{figure}

The simulated transfer functions are shown in Figure \ref{fig:cmp_trans}, with more signal loss on smaller $k$ scales.  The statistical error of the transfer function is below 3\% level, which only contributes to $\sim 1\%$ of the total variance of the cross-correlation. The transfer functions are comparable for both PCA and ICA.

Note that in a foreground removal simulation, especially for the HI auto-correlations, one may obtain more optimistic results than in the analysis of real data, because the model may be too simplistic and inadequate in its representation of the complexities of the real world foreground and instrument response.  Fortunately, the foreground residues are usually uncorrelated with the cosmological signal, so in the cross-correlation these are suppressed, and our forecast on the cross-correlation thus works better than on the auto-correlations.

\section{Power spectrum and cosmological parameters estimation}\label{sec:power}
Below we infer cosmological parameters from the mock observations described above. The optical galaxy auto-correlation power spectrum is given by
\begin{eqnarray}
     \label{eq:pk_g}
     P_g(\mathbf{k}) &=& V \langle \delta_g(\mathbf{k})\delta^*_g(\mathbf{k}) \rangle -P_\txt{shot},
 \end{eqnarray}
where $P_\txt{shot}=V/N_g=1/\langle n_g\rangle$ is the shot noise term for the galaxy survey, and
$V=(205\unit{Mpc}/h)^3$ is the box volume, while the cross-correlation power spectrum with the 21cm brightness temperature is given by
\begin{eqnarray}
     \label{eq:pk_c}
     P_X(\mathbf{k}) &=& V\Re\left(\delta_T(\mathbf{k})\delta^*_g(\mathbf{k})\right)
\end{eqnarray}
The error  for the auto-correlation of optical galaxies and the cross-correlation power spectrum are estimated as \citep{FKP1994,wolz2017}
\begin{eqnarray}
    \label{eq:gal_cosmic_var}
    \sigma_{P_g}(k) &=& \frac{2\pi}{\sqrt{V_\txt{sur}k^2\Delta k}} \left(P_g(k) + P_\txt{shot}\right),\\
    \label{eq:cross_cosmic_var}
    \sigma_{P_X}(k) &=& \frac{2\pi}{\sqrt{2V_\txt{sur}k^2\Delta k}} \times \\
    &&\sqrt{P_X^2(k) + \left(P_T(k)+P_\txt{ns}(k)\right)\left(P_g(k)+P_\txt{shot}\right)},\nonumber
\end{eqnarray}
where $\Delta k$ is k-bin width, $P_\txt{ns}(k)$ is the power spectrum of the thermal noise. 

\begin{figure}
\fig{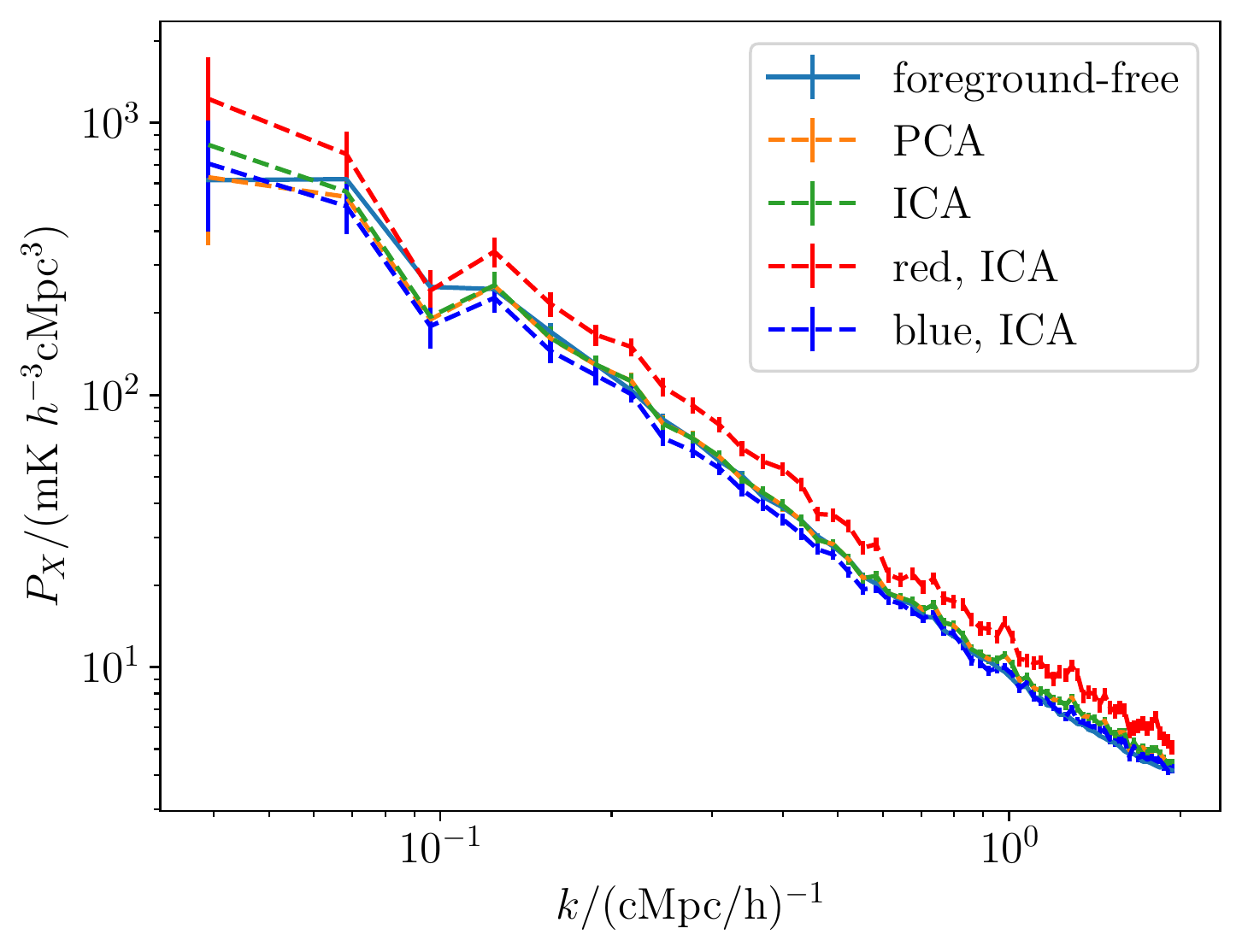}{0.45\textwidth}{}
\caption{The HI-optical galaxy cross-correlation power spectrum. We show the foreground-free (obtained directly from the simulation box, without foreground), and after foreground removal and transfer using the two methods. The HI-red galaxy and HI-blue galaxy cross power spectra are also shown, using ICA foreground removal.
}
     \label{fig:compare_method}
\end{figure}

In this paper we have used the real space position data of the simulation. In an actual observation,  the peculiar motion would induce a redshift space distortion (RSD) effect. On linear scales, the RSD introduces a correction factor of the form  $(1 + \beta\mu^2)^2$ in the power spectrum, where $\mu$ is the wave vector direction cosine with respect to the line of sight, $\beta=f/b$, with $f$ the growth factor and $b$ the bias \citep{10.1093/mnras/227.1.1}. With good anisotropic power spectrum measurement, in principle one can break the degeneracy between $b_\txt{HI}$ and $\Omega_\txt{HI}$ using the RSD effect\citep{PhysRevD.81.103527}, but we will not use this additional information here.

Although the Galactic foreground is uncorrelated with the galaxy sample, due to its much higher intensity,  its fluctuation still dominates the cross-power spectrum if it is not removed.  The foreground removal can efficiently enhance the signal-to-noise ratio in the cross-correlation power spectrum. We find that if the foreground is subtracted to a 1\% level, which is still much stronger than the 21 cm signal, the cross-correlation power spectrum can already be measured pretty well, with the signal dominating over the foreground contribution.  In Fig.~\ref{fig:compare_method} we plot the mock observation cross power spectrum for the snap84, as well as the foreground-free spectrum. Different foreground subtraction techniques are also compared, though in this simulation there is no substantial difference. All results are quite consistent with each other. Hereafter, we only discuss the result using ICA for foreground removal.

As discussed earlier and shown in Fig.~\ref{fig:CMD_absem}, here we can also divide the galaxy sample into ``red" ($g-r>0.6$) and ``blue" ($g-r\leq 0.6$) ones.  The HI-red galaxy and HI-blue galaxy cross power spectra are shown in Fig.~\ref{fig:compare_method}. 
The cross power spectrum for HI-red galaxy is slightly higher than the blue ones, which is not unexpected as the red galaxies have a larger bias factor, as have been found previously (see e.g. \citet{2002ApJ...571..172Z,10.1111/j.1365-2966.2005.09457.x,10.1111/j.1365-2966.2008.14082.x}). However, as we will show later the blue galaxies are actually more correlated with the neutral hydrogen than the red galaxies, which is consistent with previous works (e.g.\citet{2020MNRAS.494.2664D,wolz2021hi}, as blue galaxies are usually more HI-rich than red galaxies.

\begin{figure}
\gridline{\fig{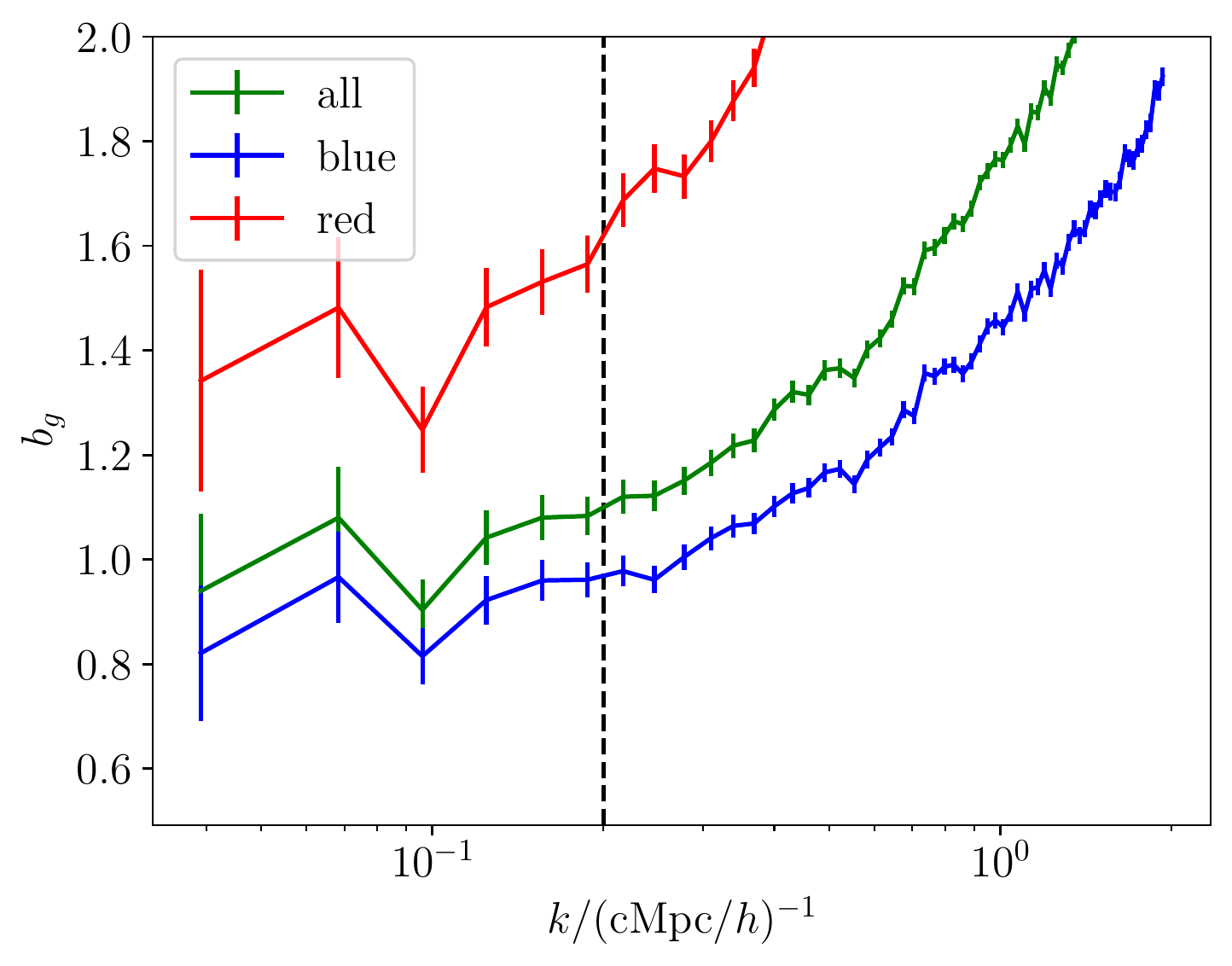}{0.45\textwidth}{}
               }
\caption{The galaxy bias factor $b_g$ for all optical galaxies, red galaxies and blue galaxies in snap84.  The vertical dashed line is $k=0.2\ ({\rm Mpc}/h)^{-1}$. 
     \label{fig:bias}}
 \end{figure}

The optical galaxy auto-correlation power spectrum is related to the matter power spectrum as \citep{wolz2017}
\begin{eqnarray}
     P_g(k) &=&  b_g^2 P_\txt{lin}(k),
\end{eqnarray}
 $b_g$ is the bias of galaxy,  $P_\txt{lin}(k)$ is the linear matter power spectrum,
while the 21cm auto-correlation power spectrum is (\citet{wolz2017}) 
\begin{eqnarray}
     P_T(k) &=& T_0^2 \Omega_{\rm HI}^2 b_{\rm HI}^2 P_\txt{lin}(k),
\end{eqnarray}
$\Omega_{\rm HI}$ is the density parameter for HI, $b_{\rm HI}$ is the bias of HI.
The cross-correlation power spectrum is \citep{wolz2017}
\begin{eqnarray}
     P_X(k) &=& T_0 \Omega_{\rm HI} b_{\rm HI} b_g r_{\txt{HI}, g} P_\txt{lin}(k),
\end{eqnarray}
where we introduced the galaxy-HI correlation coefficient $r_{\txt{HI}, g}$, to account for the fact that the optical and radio traces may not be perfectly correlated.

\begin{deluxetable}{cll}
     \tablecaption{$b_g$ for all galaxies, blue galaxies and red galaxies for snap84 and snap78. 
     }\label{tab:gal_bias}
     \tablewidth{0pt}
     \tablehead{
     \colhead{Catalog} & \colhead{snap84 ($z=0.2$)} & \colhead{snap78($z=0.3$)}\\
     }
     \startdata
     all &  $1.040\pm 0.022$ & $1.036\pm 0.022$ \\
     blue & $0.926\pm 0.020$ & $0.963\pm 0.021$\\
     red &  $1.471\pm 0.032$ & $1.508\pm 0.034$\\
     \enddata
 \end{deluxetable}

Given the matter power spectrum, the bias factor $b_g$ can be measured from the optical galaxy auto-correlation power spectrum. We calculate the linear matter power spectrum by using CAMB \citep{lewis2000efficient}. 
The cosmological parameters are set to be consistent with those assumed in the  IllustrisTNG simulation.

We plot the bias factor $b_g$ in Fig.~\ref{fig:bias}. 
As we can see, the bias factor on small scales (large $k$) varies with scale, due to the non-linear evolution which is more prominent on small scales. The collapse of matter into halos generates extra power at small scales compared with the linear theory and thus boosts up the small scale power spectrum, but on large scales the bias approaches a constant. 
We can infer the bias from the power spectrum with $k< 0.2\ ({\rm Mpc}/h)^{-1}$ for the whole sample, and for the blue,\ and red galaxy sub-samples. This scale is marked by the dashed vertical lines in Fig.~\ref{fig:bias}. 

The average value and standard deviation of the posterior distribution of the bias factors are given in Table~\ref{tab:gal_bias}, the values are comparable with the value in literature, e.g. \citet{10.1111/j.1365-2966.2008.14082.x}.
For the whole catalog with all galaxies $b_g\sim 1$, the red galaxies have a bias of $b_g \sim 1.4$, while the blue galaxies have a bias value of $b_g \sim 0.9$. The bias also decreases slightly from snap78 centered at $z=0.3$ to snap84 centered at $z=0.2$.

\begin{figure}
\centering
\includegraphics[width=0.45\textwidth]{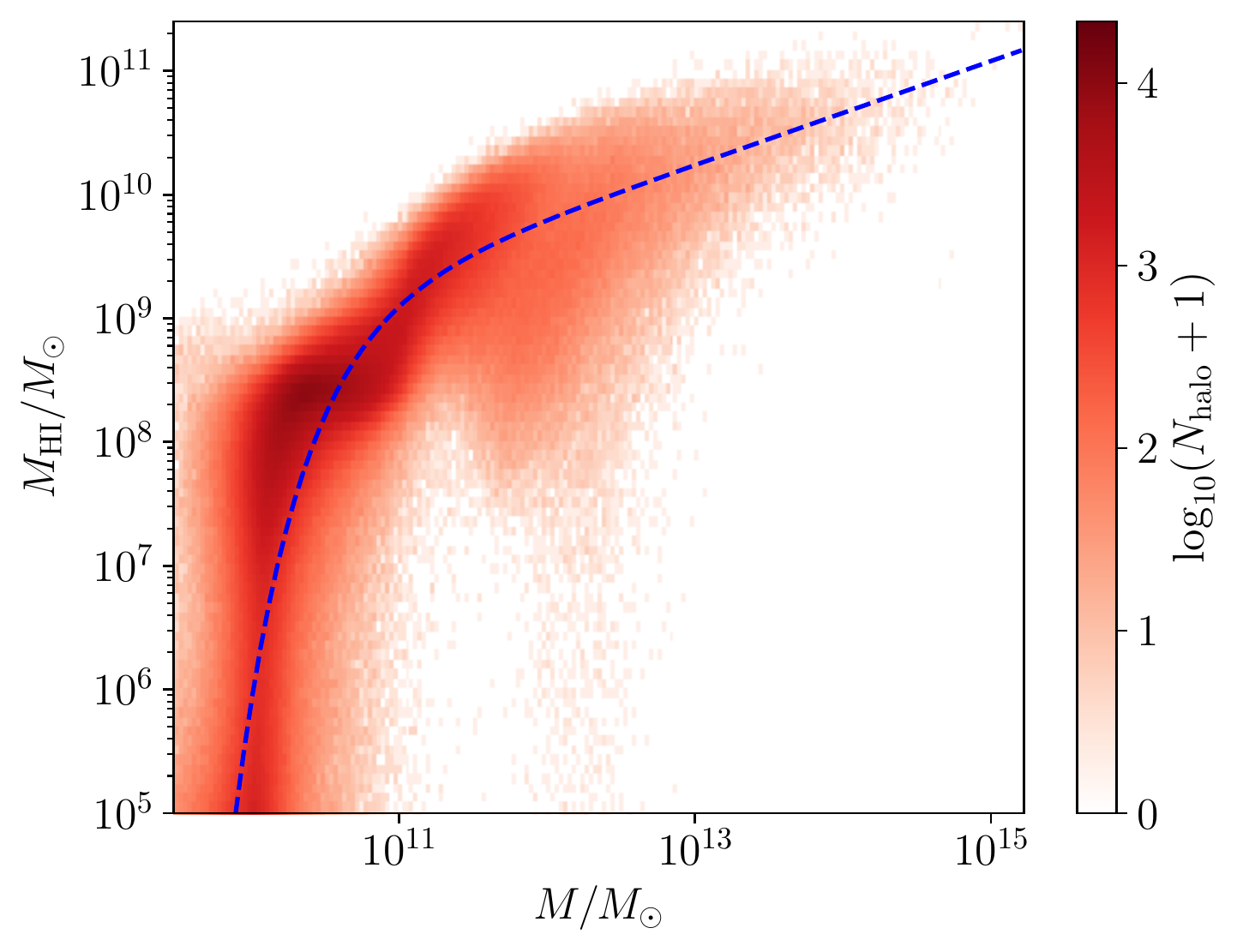}
\caption{HI mass as a function of halo mass, the dash curve shows the result of fitting $M_{\rm HI}(M)$ using Eq.~(\ref{eq:MHI}) for snap84}
 \label{fig:halofit}
 \end{figure}  
 
\begin{deluxetable}{ccc}
     \tablecaption{The HI-halo mass fitting function, HI bias and HI abundance.  
     }\label{tab:MHI}
     \tablehead{
     \colhead{Snapshot} & \colhead{snap84 ($z=0.2$)} & \colhead{snap78($z=0.3$)}\\
     }
     \startdata
    $M_0$ & $2.2\times 10^9 M_\odot$ & $1.8 \times 10^9 M_\odot$ \\
    $M_{\rm min}$ & $7.2\times10^{10} M_\odot$ & $6.3\times10^{10} M_\odot$\\
    $\alpha$ & 0.42 & 0.45\\
    $b_{\rm HI}$ & 0.839 & 0.870\\
    $\Omega_{\rm HI}$ & $5.20\times 10^{-4}$ & $5.07 \times 10^{-4}$\\
     \enddata
 \end{deluxetable}

The $b_\txt{HI}$ can in principle be obtained from 21cm auto-correlation spectrum $P_T(k)$, if it is successfully measured in the intensity mapping survey. However, as we noted earlier, it is more difficult to measure than the cross-correlation power spectrum. Here we do not assume that it is available. Alternatively, it can be computed if the halo HI mass function $M_\txt{HI}(M)$ can be obtained through simulations and observations of HI galaxies. 
Following \citet{2017PhRvD..96f3525L}, we estimate the corresponding HI bias: 
\begin{eqnarray}
     b_\txt{HI} &=& 
     \frac{\int_{M_\txt{min}}^{M_\txt{max}}dM\frac{dn}{dM}(M,\,z)M_\txt{HI}(M)b(M,\,z)}
     {\int_{M_\txt{min}}^{M_\txt{max}}dM\frac{dn}{dM}(M,\,z)M_\txt{HI}(M)}, 
\end{eqnarray}
where $dn/dM$ is the halo mass function,
for which we use the fitting function given by \citet{tinker2008toward}, with $M_\txt{min}=10^{8}M_\odot/h$ and $M_\txt{max}=10^{17}M_\odot/h$. The halo mass is defined as the total mass within a sphere whose mean density is 200 times the mean density of the Universe at the time of the snapshot. 
For $b(M)$, we adopt the fitting formula from \citet{Tinker2010}, which is in good consistency with TNG300 simulation (\citet{springel2017}).

We plot the HI mass and the halo mass of our simulation in Fig.~\ref{fig:halofit}. There is a large spread in the HI-halo mass relation, but we can see there is a clear trend. There is also a clear break at the halo mass of about $7 \times 10^{10} M_\odot$. 
We fit the $M_\txt{HI}(M)$ in simulation with following formula,
\begin{eqnarray}\label{eq:MHI}
    M_{\rm HI}(M) &=& M_0\left(\frac{M}{M_{\rm min}}\right)^\alpha\exp\left(-\frac{M_{\rm min}}{M}\right)
\end{eqnarray}
We only fit halos with mass larger than $3\times 10^9\ M_\odot$, which contain more than 99\% HI mass. The fitting parameters and HI bias $b_\txt{HI}$ are given in Table \ref{tab:MHI}. The HI abundance is also measured directly from the simulation box, and listed in Table \ref{tab:MHI}.
Our result is consistent with \citet{2018ApJ...866..135V} for $\alpha$ and $M_\txt{min}$, which are the relevant parameters for bias calculation. 
We also measured $b_\txt{HI}$ from the auto-correlation power spectrum of HI, obtained directly from the simulation box. The result is in good agreement with the value in Table~\ref{tab:MHI}, and the relative error is within 3\%. 

\begin{figure}[!ht]
\centering
\includegraphics[width=0.45\textwidth]{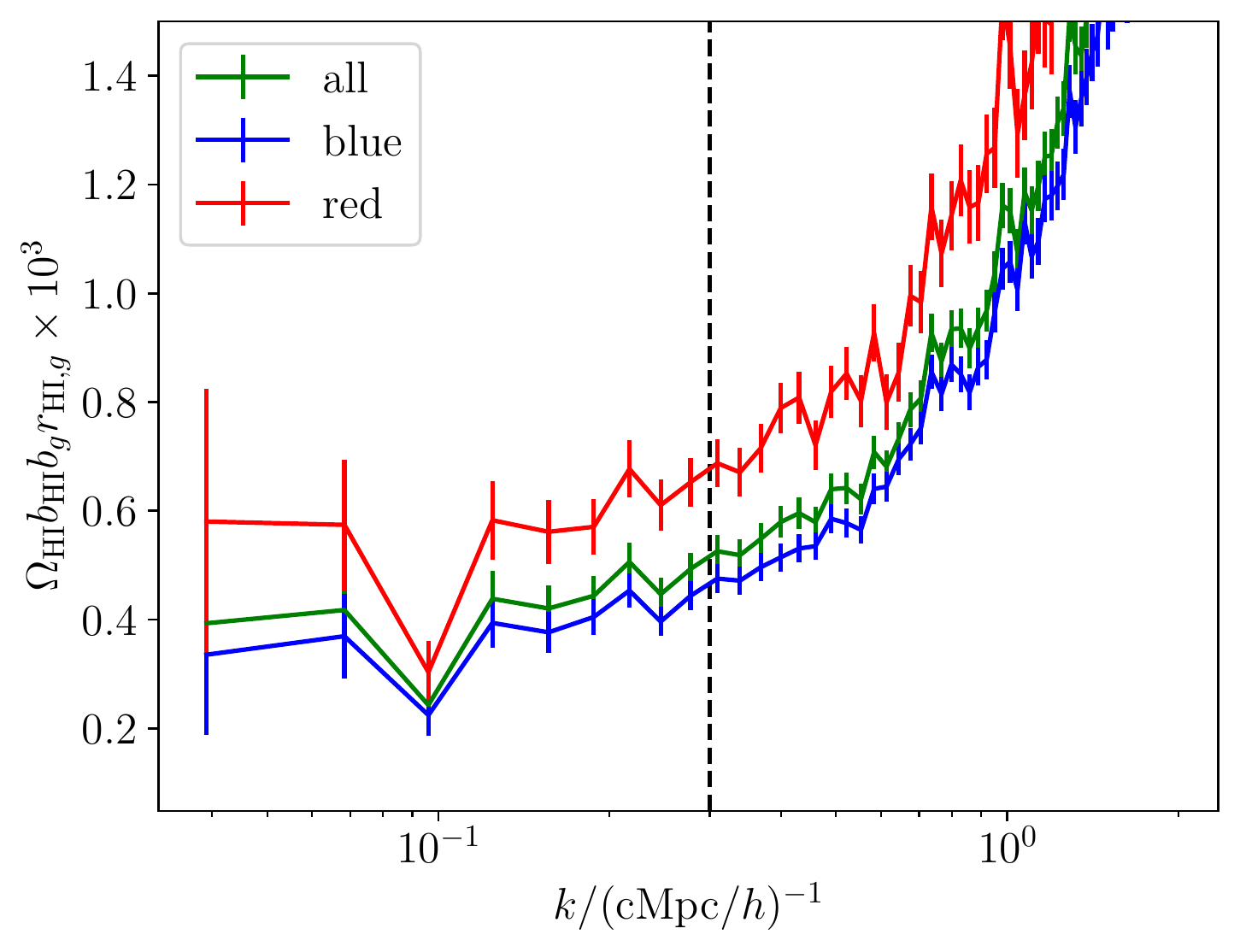}
\caption{The $\Omega_{\rm HI} b_{\rm HI} b_g r_{\txt{HI}, g}$  for snap84. The vertical dashed line is $k=0.3\ ({\rm Mpc}/h)^{-1}$.
\label{fig:bias_cgk}}
\end{figure}

\begin{deluxetable*}{ccccccccccc}
\tablecaption{The measured value and error($1\sigma$) of $10^3\Omega_{\rm HI} r_{\txt{HI}, g}$ for all galaxies, as well as ``blue" and ``red" galaxies. We also list the thermal noise $\sigma_{\rm ns-re}$ and cosmic variance $\sigma_c$.
}\label{tab:cross_bias}
\tablehead{
\colhead{Catalog} & \multicolumn{5}{c}{snap84} & \multicolumn{5}{c}{snap78}\\
\colhead{} & \colhead{Input} & \colhead{Value\,$\pm$\,Error} & \colhead{$\sigma_{\rm ns-re}$} & \colhead{$\sigma_c$}  & \colhead{$r_{\txt{HI}, g}$} & \colhead{Input} & \colhead{Value\,$\pm$\,Error} & \colhead{$\sigma_{\rm ns-re}$} & \colhead{$\sigma_c$} & \colhead{$r_{\txt{HI}, g}$}\\
}
\startdata
all  & 0.516 & $0.512\pm 0.016$ & 0.008 & 0.012 & 0.977 & 0.526 & $0.523\pm 0.018$ & 0.014 & 0.013 & 0.990\\
blue & 0.521 & $0.518\pm 0.016$ & 0.008 & 0.012 & 0.988  & 0.526 & $0.523\pm 0.018$ & 0.014 & 0.013 & 0.996\\
red  & 0.482 & $0.477\pm 0.016$ & 0.009 & 0.013 & 0.920 & 0.506 & $0.502\pm 0.019$ & 0.015 & 0.013 & 0.930\\
\enddata
\end{deluxetable*}

Finally we can proceed to infer the total neutral hydrogen abundance, from the cross-correlation power spectrum. 
In Fig.~\ref{fig:bias_cgk} we plot the quantity  $\Omega_{\rm HI} b_{\rm HI} b_g r_{\txt{HI}, g}$ as derived from the cross-correlation. We measure the bias at $k \le 0.3 ({\rm Mpc}/h)$ as marked by the dashed vertical line in Fig.~\ref{fig:bias_cgk}. With $b_g$ determined from the optical galaxy auto-correlation, we can then infer $\Omega_{\rm HI} b_{\rm HI} r_{\txt{HI}, g}$. 
Hereafter we assume that the $M_{\rm HI}(M)$ has already be determined by other observations except a normalization factor $M_0$ and use the theoretical $b_\txt{HI}$ in Table.~\ref{tab:MHI} for parameter estimation. Then the $\Omega_{\rm HI} r_{\txt{HI}, g}$ can be constrained.

The inferred value of $\Omega_{\rm HI} r_{\txt{HI}, g}$ for the three catalog are given in Table~\ref{tab:cross_bias}. 
The central value is calculated by averaging over different realizations of noise to suppress the statistic error. The error given is estimated for the volume of the simulation box.
The inferred value is slightly smaller than the input value, which might be due to the foreground removal procedure. Nevertheless, we note that the difference between the input and inferred values is within the estimated total error. We find that including two more small scale (high $k$) modes can reduce the difference, because smaller scale modes are more robust against foreground removal and also contribute more to the final result due to lower error. However, these modes may be mildly non-linear.

The factor $r_{\txt{HI},g}$ quantifies how the HI are correlated with the galaxies as found in the optical survey. In our model, it should be fairly close to 1 at large scale.
To check this, we calculate $r_{\txt{HI}, g}$ by averaging the quotient $P_X(k)/\sqrt{P_g(k) P_T(k)}$ with $k<0.2\ (\txt{Mpc}/h)^{-1}$. 
The result is given in Table~\ref{tab:cross_bias}, and we find that the $r_{\txt{HI},g}$ values are indeed quite close to 1, especially for the `blue' catalog. 
Thus if we have a robust estimation of halo HI mass function $M_\txt{HI}(M)$, we should be able to infer $\Omega_\txt{HI}$ robustly from the cross-correlation result. 
Although in our analysis, $r_{\txt{HI}, g}<1$ for all cases, we notify that it can be larger than 1 if the shot noise of optic galaxies dominates over $P_g(k)$. See \citet{wolz2017} for more details.

We estimated the contribution of thermal noise, by generating 100 different realizations of noise, but with the same input foreground and 21 cm map. We calculated the Bessel-corrected (i.e. dividing by $n-1$) standard deviation between the inferred value of different realizations. The error obtained this way is given as $\sigma_{\rm ns-re}$ (i.e. noise realization) in Table \ref{tab:cross_bias}.
According to Eq.~(\ref{eq:cross_cosmic_var}), the error in the cross-correlation power spectrum comes from the errors in both the 21cm power spectrum and the optical galaxy power spectrum. The error in 21cm power spectrum is dominated by thermal noise. 
We find that $\sigma_{\rm ns-re}$ can be represented quite well by the thermal noise term in Equation (\ref{eq:cross_cosmic_var}), i.e. 
\begin{equation}
\sigma_{\rm ns-re} \approx \frac{2\pi}{\sqrt{2V_\txt{sur}k^2\Delta k}} \frac{\sqrt{ P_\txt{ns}(k)[P_g(k)+P_\txt{shot}]}}{P_\txt{lin}(k) b_g b_{\rm HI} T_0}.
\label{eq:ns-re}
\end{equation}
The impact of the thermal noise is larger for snap78, because a given comoving size corresponds to a smaller angular size at higher redshift, so the effective observation time for each pixel is smaller at higher redshift, and the thermal noise level is higher. Here we assumed the same system temperature, in fact the system temperature could also be slightly higher at higher redshifts, due to the increasing sky temperature at lower frequencies (c.f. Eq.~\ref{eq:ns_sig}). 
If we observe a sky area for $N_\txt{scan}$ times, the thermal noise can be reduced as  $\sigma_{\rm ns-re}\propto 1/\sqrt{N_\txt{scan}}$. Thus, if $N_\txt{scan}=4$, the thermal noise contribution will be sub-dominant for both the snap84 and snap78. 

The shot noise in the optical galaxy power spectrum is fairly small for the CSST, as it has a very large comoving number density.
We also calculated the cosmic variance without thermal noise contribution for the simulation box size, which is given as the $\sigma_c$ column of Table~\ref{tab:cross_bias} by scaling the posterior error of $\Omega_{\rm HI} b_{\rm HI} b_g r_{\txt{HI}, g}$ by a factor of $b_g b_\txt{HI}$. This is quite large as our simulation box size is limited, but for the large CSST survey we are considering, the cosmic variance would be much reduced, as we shall discuss below. 
We do not propagate the cosmic variance of $b_g$ into the final error, because the galaxy and HI share the same cosmological realization and therefore their cosmic variance are correlated. The error given is statistical, the modeling error for $b_{\rm HI}$ is not included.

We now extend our result by considering the realistic survey with a larger sky area. We set the sky area for the intensity mapping experiment as $10000\unit{deg^2}$. The frequency range is set as 90 MHz around $z=0.2$ (1141 to 1231 MHz) and $z=0.3$ (1049 to 1139 MHz). The total survey volume is then $V_\txt{sur}\approx 2.45\times 10^8\ (\txt{Mpc}/h)^3$ for $z\approx 0.2$ and $V_\txt{sur}\approx 5.83\times 10^8\ (\txt{Mpc}/h)^3$. The sky area of CSS-OS is $17500\unit{deg^2}$, though for cross-correlation the area is to be set as the smaller of the surveys, i.e. $10000\unit{deg^2}$. The redshift range of CSS-OS is set the same as intensity mapping.

 \begin{deluxetable}{lll}
     \tablecaption{The value of $10^3\Omega_{\rm HI} r_{\txt{HI}, g}$ and measurement error obtained from HI-optical galaxy cross-correlation for $10000\unit{deg^2}$ sky area. }\label{tab:hi_bias}
     \tablehead{
     \colhead{Catalog} & \colhead{snap84} & \colhead{snap78}
     }
     \startdata
     all &  $0.5118\pm 0.0029$ & $0.5226\pm 0.0022$\\
     blue & $0.5176\pm 0.0029$ & $0.5231\pm 0.0022$\\
     red &  $0.4770\pm 0.0030$ & $0.5020\pm 0.0024$\\
     \enddata
 \end{deluxetable}

The inferred value of $\Omega_{\rm HI} b_{\rm HI} r_{\txt{HI}, g}$ and its error is given in Table \ref{tab:hi_bias} by scaling errors in Table.~\ref{tab:cross_bias} by a factor of $\sqrt{V_\txt{box}/V_\txt{sur}}$. Here the error is the total, including both thermal noise and cosmic variance, and the cosmic variance has been scaled to that of the survey area of $10,000 \unit{deg}^2$.
For this much larger volume, the cosmic variance is significantly reduced, and the errors are nearly the same for the blue, red, and all galaxies samples. 

Comparing these results with the existing cross-correlation measurements using GBT HI IM survey data and a number of optical surveys given in Table~\ref{tab:hi_bias_compare}, we see that a much better precision in the measurement can be achieved with the FAST HI survey and CSST survey.
The improvement is more than an order of magnitude compared with the current best result, the relative error of $\Omega_{\rm HI}$ can be suppressed to a level of $\sim 0.6\%$.

\section{Conclusion and Discussion}\label{sec:conclusion}

The CSST is a major space astronomy project in the coming decade, and the CSS-OS is the primary sky survey planned for the CSST, with a unprecedented area and magnitude limits for optical surveys.  In this work, we consider the potential of the cross-correlation of the CSS-OS spectroscopic survey with the HI survey of the FAST. The HI survey has generally been challenging, as the signal is relatively weak and prone to foreground contamination. The cross-correlation with an optical survey will greatly enhance the signal, making the detection and measurement much more accessible. On the other hand, the HI observation will reveal the distribution of the neutral hydrogen, the basic construction material of the galaxies. The cross-correlated observation would provide a solid check on our basic understanding of the large scale structure and working of galaxy formation and evolution, and opportunities for finding deep and hidden relations between the gas, stars and dark matter.

We generate realistic samples of galaxy catalogs and HI maps from the IllustrisTNG300 simulation, and present the construction of these mocks in detail. For the optical mock data, we estimate the SNR of the CSS spectrometer for each galaxy in the simulation, to produce an SNR-limited sample. The sample is naturally divided into red and blue galaxies by their color. For the HI mock data, we have taken into account the noise, a simple beam model, and galactic foreground. We focus on redshift $0.2\sim 0.3$, which is covered by the L-band multi-beam receiver in the FAST. 

Based on the mock data, we simulated foreground removal with the PCA and ICA methods, which in the simulation produce similar results. We produce foreground-removed sky maps, to cross-correlate with the optical sample, and derive the power spectrum. Finally, we consider using this observation to measure cosmological parameters. The HI-optical cross-correlation is particularly suitable for measuring the total HI content at the low redshift Universe. We assume the bias factor $b_g$ can be obtained from the auto-correlation of the optical galaxies, and the HI-optical cross-correlation measures $\Omega_{\rm HI}b_{\rm HI}r_{\txt{HI}, g}$, where $b_{\rm HI}$ is the HI bias, and $r_{\txt{HI}, g}$ is the correlation factor for the HI and optical galaxies. We can then infer the value of $\Omega_{\rm HI}$. Furthermore, by cross-correlating with galaxies of different colors, the HI content of these can be separately measured. 

In the present work we have ignored the light cone effect, i.e. the evolution of galaxy statistics from the redshift corresponding to the far side of the simulation box, to that of the near side. However, the estimated error is not too much affected. We also limited our discussion mainly to the measurement of $\Omega_{\rm HI}$. The power spectrum features such as the BAO wiggles can also be used to constrain other cosmological parameters, though the optical galaxy spectrum would already provide a very good constraint on that.

With the high comoving number density that can be achieved by the CSST survey, the shot noise in the optical galaxy power spectrum is much suppressed. The cosmic variance is also anticipated to be very small for the large overlapping area of the CSST and FAST surveys.
A very accurate measurement of $\Omega_{\rm HI}$ is expected, with a relative error at the $0.6\%$
level. This illustrates the power of the upcoming CSST survey.

\section*{Acknowledgement}\label{sec:acknowledgement}
We thank Qi Guo, Haitao Miao, Hu Zou and Xin Zhang (NAOC) for the discussions. FD and XC are supported by the Ministry of Science and Technology (MoST) inter-government cooperation program China-South Africa Cooperation Flagship project 2018YFE0120800, the China Manned Space Project with grant No. CMS-CSST-2021-B01. YG acknowledges the support of MOST-2020SKA0110402, NSFC-11822305, NSFC-11773031, NSFC-11633004, and CMS-CSST-2021-A01. YW acknowledges the 
the National Key R\&D Program 2017YFA0402603, and the CAS Interdisciplinary Innovation Team (JCTD-2019-05). SD acknowledge Chinese Academy of Sciences College Student Innovation Training Program (No. 20214000231). 

\section*{Data Availability Statement}
The data underlying this article will be shared on reasonable request to the corresponding author.

\bibliography{cross}{}
\bibliographystyle{aasjournal}

\end{document}